\documentstyle[preprint,tighten,eqsecnum,aps]{revtex}

\begin{document}
\draft
\preprint{\parbox[t]{2 in}{CMU-HEP93-04\\
DOE/ER/40682-29}}

\title{Isospin splitting in heavy baryons and mesons}

\author{R. E. Cutkosky and Paul Geiger}
\address{Physics Department, Carnegie Mellon University, Pittsburgh,
PA 15213\\}

\maketitle

\begin{abstract}
   A recent general analysis of light-baryon isospin splittings
is updated and extended to charmed baryons.
   The measured $\Sigma_c$ and $\Xi_c$ splittings stand out
as being difficult to understand in terms of two-body forces
alone.
   We also discuss heavy-light mesons; though the framework here is
necessarily less general, we nevertheless obtain some predictions
that are not strongly model-dependent.
\end{abstract}
\pacs{13.40.Dk, 12.40.Aa, 14.20.-c, 12.70.+q}

\narrowtext

\section{Baryons}
\label{sec:Baryons}

   Mass splittings among baryons and mesons containing different
valence quarks
provide a good probe of hadronic structure and of the forces
that determine this structure~\cite{GeneralRefs}.
   Recently, results were reported for a general class of models in which
the baryon octet and decuplet splittings were assumed to arise solely
from two-body interactions~\cite{REC92}.
   It was found that the central masses of isospin multiplets
could be fitted with a model error of 8.9 MeV.
A similar parametrization of isospin splittings,
also with a small model error, showed that the two measured
$\Delta$ masses were the hardest to fit.

   In this paper, we shall extend the analysis to baryons
containing heavy quarks (thereby severely testing the assumption that
three-body forces are negligible), and also to mesons.

   Since reporting on the fit in Ref.~\cite{REC92} we have recognized
that the two sets of input $\Delta$ values that had been averaged\cite{KPA}
were obtained using essentially the same set of input experimental
data.  Very similar analysis methods were also used, so the differences in the
results should be treated as a systematic error.  In addition, we have
concluded, through discussions with colleagues\cite{GHRA}, that there
are additional systematic errors arising from differences in the
treatment of the coupling to open channels.  We estimate these
theoretical uncertainties to be, at this time, about 0.5 MeV for the
$\Delta $ states.  Similar systematic uncertainties would also occur
in $\Sigma^*$ and $\Xi^*$ states.   Since their widths are smaller and
the errors are larger, we
expect that
this would give a negligible additional
uncertainty to the masses of these states.  With the enlarged error
for the $\Delta$ masses, the fit to octet and decuplet masses
has a $\chi^2$ of 3.2 for 4 degrees of freedom,
so that no model error needs to be invoked.
However, almost all of this
$\chi^2$ still arises from the $\Delta$ masses. The results from this
modified fit are shown in Fig.~\ref{isospin}.  The
parameters from this fit will be used in fitting the heavy quark
sector.

Although a breakdown of the spectator approximation
in the charmed baryon sector would not be at all
surprising (since the heavy charm quark
may significantly alter the environment in which the light quarks
interact), we will in fact see that there appears
to be little indication of such a breakdown in most of the data.

    The central masses of the charmed baryons may be expressed
in terms of four pair energies $S_{nc}$, $T_{nc}$, $S_{sc}$, and $T_{sc}$
(in addition to the non-charm pair energies introduced in
Ref.~\cite{REC92})
or in terms of the spin-dependent and spin-independent combinations
$D \equiv (T - S)/4$ and $A \equiv (3T + S)/4$, respectively.
($T$ indicates a spin-triplet quark pair, $S$ a spin-singlet quark
pair, and the subscripts give the flavor content.)
   However, there are only four data, so we introduce an additional
assumption.  In Ref.~\cite{REC92} it was verified that
$x=D_{ns}/D_{nn} \sim \frac{2}{3}$, as is often assumed in models.
Here we also assume that $x_c=D_{sc}/D_{nc} \sim \frac{2}{3}$.
   To give this constraint some flexibility, we
introduce the auxiliary quantity $y_c =\frac{3}{2}D_{sc}-D_{nc}$, with
an input value of $0.0 \pm 1.6$ MeV.
    Tables~\ref{tab:CharmCentral1} and \ref{tab:CharmCentral2}
summarize the fit of these pair energies to the $\Lambda_c$,
$\Sigma_c$, $\Xi_c$, and $\Omega_c$ masses\cite{OmegaC}.
Including the correlated parameter errors from the non-charm sector
shown in Table~\ref{tab:CharmCentral2} and the same model error of
$8.9$ MeV, we find an overall $\chi^2$ of 0.37
with one degree of freedom, thus the central masses are
easily fitted with a sum of two-body energies.

    We may also express the central masses of the remaining (unmeasured)
singly-charmed baryons in terms of pair energies to obtain the
predictions shown in the left side of Table~\ref{tab:CharmPreds}.
(Note that there are two $\Xi_c$ states, since there is no
symmetry restriction on the spin wavefunction of a $usc$ baryon.
We denote the $\chi^{\rho}$ state by $\Xi_c$ and the
$\chi^{\lambda}$ state by $\Xi_c^{\prime}$. These states
actually mix slightly --- and this was taken into account
in our calculations --- but the effects are negligible.
(See also Ref.~\cite{XiPrime}.)

  The fits and predictions for $j = \frac{1}{2}$ and $j = \frac{3}{2}$
baryons are shown in Fig.~\ref{central}.
  For the fits, the quantities plotted are the contributions of the
spin-dependent terms $D$.  The spin-independent terms $A$ have been
subtracted from the experimental data. For the non-charm baryons, we
assume that $D_{ss}=D^2_{ns}/D_{nn}$.

   We have also considered an additional stronger assumption, that the
$A$'s are linear in the quark masses.  In the non-charm sector, the
three $A$'s are then fitted with two parameters, with $\chi^2=1.86$,
which is not excessive.  In the charm sector, however, fitting the two
new $A$'s with one new parameter gives $\chi^2=24.2$.  This indicates
that the presence of the charm quark does affect the baryon structure,
even though the fit to central masses shows no clear evidence for
the existence of three-body effects.

   The fit to isospin splittings proceeds similarly. There is no
advantage in discussing the individual $\Sigma_c$ and $\Xi_c$ charge
states, because their experimental errors are correlated in a
complicated way.  We therefore give results for the isospin tensorial
quantities
\begin{mathletters}
\label{eq:charmB}
\begin{eqnarray}
\label{eq:Sigma1}
\Sigma_c^1 &=& \Sigma_c^{++} - \Sigma_c^0 \; , \\
\label{eq:Sigma2}
\Sigma_c^2 &=& \Sigma_c^{++} + \Sigma_c^0 - 2 \Sigma_c^+ \; , \\
\label{eq:Xi1}
\Xi_c^1 &=& \Xi_c^+ - \Xi_c^0 \; .
\end{eqnarray}
\end{mathletters}
The new pair terms, $D^1_c$ and $A^1_c$ are shown in
Table~\ref{tab:CharmIso1}.    Note that $D^1_c$ is very small and has
little influence on the splittings; rather than treat it as a free
parameter, we constrained it in the fit to $(0.03 \pm 0.05)$ MeV, an
estimate obtained from the sizes and uncertainties of various terms in
the simple model described in Ref.~\cite{REC92}.  The best-fit value
was $D^1_c=0.027$ MeV.  The total $\chi^2$ for the fit is $6.23$.
   The final two rows of the table are clearly the most intriguing:
We obtain a poor fit to $\Xi_c^1$ and a worse fit to
$\Sigma_c^2$. If these measurements are correct, it appears that
three-body effects are important for isospin splittings but
not for the isospin-averaged central masses.

The errors in the non-charm pair terms are correlated; our fit uses
the full correlation matrix, and the shifts in the values of these
terms are shown in Table~\ref{tab:CharmIso2}.  The fit to
the individual $\Sigma_c$ and $\Xi_c$ masses gives slightly adjusted
values for the central $\Sigma_c$ and $\Xi_c$ masses, which were used
as the input in Table~\ref{tab:CharmCentral1}.

As before, the pair term analysis gives rise to some predictions.
These are shown in the right side of Table~\ref{tab:CharmPreds}.
   It is interesting that the combination
$2\Xi_c^{\prime 1} - \Sigma_c^1 = (-1.88 \pm 0.51)$ MeV is given
by $-T^1 + 2T^1_s$, {\it i.e.}, it is independent of the
charm pair terms.  Note also that in our model,
$\Sigma_c^{*2}=\Sigma_c^2=\Sigma^2=\Sigma^{*2}$.

\section{Mesons}
\label{sec:Mesons}
  A description of mesonic splittings in terms of
pair energies is not useful, since a new parameter must
be introduced for each meson.
   While we are thus forced to commit ourselves to
stronger assumptions here, we still strive for results
which are as model-independent as possible.
   We focus on the {\it differential}
vector-pseudoscalar isospin splittings, which we denote
by $\Delta(M)$. In particular, we define
\begin{mathletters}
  \label{eq:Deltas}
\begin{eqnarray}
\label{eq:DeltaB}
  \Delta(B) \;& = &\; (B^{*-}-\bar{B}^{*0}) - (B^{-}-\bar{B}^{0}) \\
\label{eq:DeltaC}
  \Delta(D) \;& = &\; (D^{*0}-    {D}^{*+}) - (D^{0}-    {D}^{+}) \\
\label{eq:DeltaK}
  \Delta(K) \;& = &\; (K^{*-}-\bar{K}^{*0}) - (K^{-}-\bar{K}^{0}).
\end{eqnarray}
\end{mathletters}
   These combinations, which have been considered previously by
a number of authors\cite{DeltaAuthors}, are interesting for several
reasons:

   {\it i)} $\Delta(D)$ has recently been measured very precisely:
$\Delta(D) = 1.48 \pm 0.1$ MeV (Ref.~\cite{DeltaDmeas}). This datum
constitutes a severe test (or useful calibration point)
for model calculations.

   {\it ii)} $\Delta(K)$ is also known rather precisely:
$\Delta(K) = -0.49 \pm 0.37$ MeV (Ref.~\cite{PDG}).
Theorists have generally found it difficult to
accomodate this value; even those models which
successfully account for related splittings
typically predict
$\Delta(K) \approx +3$ MeV (see Ref.~\cite{DeltaKtheor}).
It has been suggested that relativistic and/or unitarity
effects (from $K^* \rightarrow K \pi \rightarrow K^*$,
for example) may be a source of the discrepancy.

    {\it iii)} $B^- - \bar{B}^0$ is known to be $-0.12 \pm 0.58$ MeV
(Ref.~\cite{PDG}), but the $B^*$ isospin splitting has not yet been
measured, thus there is room for a prediction of $\Delta(B)$.

    {\it iv)} As we shall presently see, it is possible to obtain
predictions for the ratios $\Delta(B):\Delta(D):\Delta(K)$
which are quite model-independent.

    We write the quark-mass and electromagnetic
contributions to $\Delta(M)$ as $\Delta^m(M)$ and
$\Delta^{\gamma}(M)$, respectively.
To obtain $\Delta^m(M)$, we employ an interpolation technique
first suggested by Chan\cite{Chan}.
    Consider, for the sake of definiteness, $\Delta^m(D)$.
{}From equation (\ref{eq:DeltaC}), we see that
$\Delta^m(D) \approx (m_u-m_d)(dE_{hyp}/dm_q)\vert _{m_q=m_u}$.
    This latter quantity may be accurately estimated with very
little reliance on theory, since the function $E_{hyp}(m_q)$
is rather well-determined by measurements of
$D^*-D$,\ $D_s^*-D_s$ and $J/\Psi-\eta_c \;$;
an interpolating curve drawn through these points
allows the extraction of
$(dE_{hyp}/ dm_q)\vert _{m_q=m_u}$~\cite{ValenceCaveat}.
    The same procedure can be applied to $\Delta^m(B)$ and
$\Delta^m(K)$, and the relevant curves for all
three cases are shown in Fig.~\ref{fig:interp}.

     In the case of D and B mesons, the hyperfine splittings
turn out to be surprisingly independent of the lighter
quark's mass:
\begin{eqnarray}
  D^*-D \,\,&=  &    141.4 \pm 0.1\;{\rm MeV,} \nonumber \\
  D_s^*-D_s &=  &    141.5 \pm 1.9\;{\rm MeV,} \nonumber \\
  B^*-B \,\,&=  &\,\,\, 46.0 \pm 0.6\;{\rm MeV,} \nonumber \\
\label{eq:HeavyHyp}
  B_s^*-B_s &=  &\,\,\, 47.0 \pm 2.6\;{\rm MeV} \, .
\end{eqnarray}
(This is qualitatively understood in the quark model,
where hyperfine splittings vary with the quark masses as
$\vert\psi(0)\vert^2/m_1m_2$; in a linear potential,
${\vert\psi(0)\vert^2}$ is proportional to the reduced mass $\mu$,
so that $E_{hyp} \propto (m_1+m_2)^{-1}$, which becomes independent
of $m_1$ in the limit $m_2 >> m_1$.)
   Thus $\Delta^m(D)$ and $\Delta^m(B)$ are both
approximately zero.
The second column of Table~\ref{tab:MesonResults} shows the actual
results of the interpolation procedure; we performed a quadratic
fit to the three data points for D mesons, and a linear fit to the
$B^*-B$ and $B_s^*-B_s$ data points for B mesons.

   The strange meson sector differs in that $E_{hyp}$
has significant $m_q$-dependence: the available data here
are $K^*-K = 398$ MeV, $D^*_s-D_s = 142$ MeV, and
$B_s^*-B_s = 47$ MeV. A cubic spline interpolation gives
$(dE_{hyp}/ dm_q)\vert _{m_q=m_u} = -0.36$,
   however, without a data point for the $s\bar{s}$
mesons, the distance over which we must interpolate
in this case is unsettlingly large.
   The $s\bar{s}$ data point is not directly available,
of course, because annihilation forces rotate the pseudoscalar
$s\bar{s}$ meson ($\eta_s$) and its
nonstrange partner ($\eta_n$) into the physical
$\eta$ and $\eta^{\prime}$ states.
   Nevertheless, we may estimate the mass that $\eta_s$
would have in the absence of such forces by making use
of the empirical rule~\cite{V2-P2} for the difference
of like-flavor vector and pseudoscalar squared masses:
$m_V^2-m_P^2 \approx 0.56 \; {\rm GeV}^2$. This implies
$m_{\eta_s}= \sqrt{m_{\phi}^2 -0.56}$ $= 690\;{\rm MeV}$.
Independently, an equal-spacing rule should hold for
the squared masses of ideally-mixed pseudoscalars so that
$m_{\eta_s}^2 = 2m_K^2 + m_{\pi}^2$,
which gives $m_{\eta_s} = 687 \;{\rm MeV}$.
   Using this additional data point
gives $d(E_{hyp}/ dm_q)\vert _{m_q=m_u} = -0.32 \pm 0.06$,
where the error estimate reflects the
dependence of the fitted slope on our choice of constituent
quark masses~\cite{CQMasses}, and on the
value of $m_{\eta_s}$.

    Almost all model calculations are consistent with
$m_d-m_u = 5\pm 1$~MeV~\cite{md-mu}. We adopt this value
and show $\Delta^m(M)$ in the third column of
Table~\ref{tab:MesonResults}.

\vspace{.30in}
    The electromagnetic contribution to $\Delta(M)$ is given by,
\begin{equation}
  \Delta^{\gamma}(M) = \alpha_{\rm em}q_h
  \left\{
  \left\langle {1\over r} \right\rangle_0 -
  \left\langle {1\over r} \right\rangle_1
  + {8\pi\over 3} { \left| \psi (0) \right|^2 \over m_h m_n }
  \right\} \; ,
  \label {eq:DeltaGamma}
\end{equation}
where the subscript $h$ refers to the heavy-flavor
quark $(h=s,c,b)$, and the subscript $n$ refers
to the light anti-quark. The quantity
$\alpha_{\rm em} q_f \langle 1/r \rangle _{0\; (1)}$
is the Coulomb energy in the pseudoscalar (vector).
  The difference
$\langle 1/r \rangle _0 - \langle 1/r \rangle _1$
is induced by hyperfine distortion of the wavefunctions;
though it is often assumed to be negligible,
explicit potential-model calculations indicate that
this Coulomb term is only about 30\% smaller than the
magnetic term in (\ref{eq:DeltaGamma})\cite{SimpleModelFlaw}.
    (We find it most convenient to discuss $\Delta^{\gamma}(M)$
in a potential-model framework, however, our results
will turn out to be nearly independent of the
parameters --- and even the functional form --- of the
potential, hence we suspect that our conclusions are
more reliable than our method.)

  Consider calculating $\Delta^{\gamma}(M)$
in a Hamiltonian whose spin-independent part is
\begin{equation}
  H_{si} = {p^2 \over 2\mu} + br
     -{4 \over 3}{\alpha_s^0 \over r}
     \left( {b \over \mu^2} \right)^{1/3} \;.
  \label{eq:CouLinHam}
\end{equation}
This is a standard Coulomb-plus-linear Hamiltonian, with a
particular ansatz for the way in which the effective strong
coupling constant varies with the reduced mass of the system:
$\alpha_s(\mu) = \alpha_s^0 \cdot (b/\mu^2)^{1/3}$.
  We shall presently show that this form for $\alpha_s(\mu)$
gives excellent phenomenology, but for the moment we just
capitalize on the fact that it yields simple scaling laws for
the wavefunctions and energies of $H_{si}$. Since
\begin{equation}
  \mu H_{si} = {p^2 \over 2} + (\mu b)r
     -{4 \over 3}{\alpha_s^0 \over r}
     (\mu b)^{1/3}  \; ,
  \label{eq:muCouLinHam}
\end{equation}
it follows that the energies are proportional to
$(b^2/\mu)^{1/3}$, and wavefunction properties with dimension
(length)$^n$ are proportional to $(\mu b)^{-n/3}$.
   Upon adding the spin-dependent term,
\begin{equation}
   H_{hyp} = {32 \pi \over 9} \alpha_s(\mu)
   {{\bf S}_h\cdot{\bf S}_n \over m_h m_n }
   \delta^3({\bf r}) \;,
   \label{eq:Hhyp}
\end{equation}
we find the following scaling laws for the hyperfine, magnetic, and
Coulomb energies:
\begin{equation}
  E_{hyp}  \equiv {32 \pi \over 9} \alpha_s(\mu)
   { \left| \psi (0) \right|^2 \over m_h m_n }
    \;\propto\; \alpha_s(\mu) \, { \mu b \over m_h m_n }
  \label{eq:Ehyp}
\end{equation}
\begin{equation}
  E_{mag}  \equiv \alpha_{em} q_h {8 \pi \over 3}
   { \left| \psi (0) \right|^2 \over m_h m_n }
    \;\propto\; { \mu b \over m_h m_n }
  \label{eq:Emag}
\end{equation}
\widetext
\begin{equation}
  E_{coul}  \equiv \alpha_{em} q_h
  \left\{ \left\langle {1\over r} \right\rangle_0 -
  \left\langle {1\over r} \right\rangle_1 \right\}
  =
  {64\pi \over 9} {\alpha_s(\mu) \over m_h m_n }
  \sum_n { \langle 0 | \delta^3({\bf r}) | n \rangle
  \langle n | {1 \over r} | 0 \rangle
  \over E_0-E_n }
  \;\propto\; { \mu b \over m_h m_n } \; .
  \label{eq:Ecoul}
\end{equation}
Thus the Coulomb and magnetic terms scale in the
same way, and we have
\begin{equation}
  \Delta^{\gamma}(B):\Delta^{\gamma}(D):\Delta^{\gamma}(K) =
  q_b { E_{hyp}(B) \over \alpha_s(\mu_{_B}) } :
  q_c { E_{hyp}(D) \over \alpha_s(\mu_{_D}) } :
  q_s { E_{hyp}(K) \over \alpha_s(\mu_{_K}) }  \; .
  \label{eq:BDKratios}
\end{equation}

   In fact, it is easy to check that the result
(\ref{eq:BDKratios}) holds not just for the specific
potential in (\ref{eq:CouLinHam}), but for {\it any}
potential of the form
\begin{equation}
  V_{\beta}(r) = Cr^{\beta} -
  {4 \over 3}{\alpha_s^0 \over r}
  \left( { C \over \mu^{\beta +1} } \right)^{1 \over \beta +2}\;.
  \label{eq:BetaHam}
\end{equation}
(And also for
$V(r) = C \log (r/r_0) -(4/3)(\alpha_s^0 /r)(C/\mu)^{1/2}$, which
behaves in many ways like the $\beta \rightarrow 0$ limit
of $V_{\beta}(r)$; we will refer to this logarithmic potential as
$V_0(r)$.) In the general case, the ratios of electromagnetic
splittings are
\begin{equation}
  \Delta^{\gamma}(B):\Delta^{\gamma}(D):\Delta^{\gamma}(K) =
  q_b \mu_{_B}^{\epsilon} E_{hyp}(B) :
  q_c \mu_{_D}^{\epsilon} E_{hyp}(D) :
  q_s \mu_{_K}^{\epsilon} E_{hyp}(K)   \; ,
  \label{eq:BDKratios2}
\end{equation}
where $\epsilon \equiv (\beta+1)/(\beta +2)$.

\narrowtext

  With regard to the
phenomenological soundness of equation (\ref{eq:BetaHam}),
Fig.~\ref{fig:AlphaS} shows the ``running'' of
$\alpha_s(\mu)$ for three cases of interest, namely $\beta =0,1,2$
(corresponding to logarithmic, linear, and harmonic
confinement, respectively).
    In each case, the curve lies close to the
empirical values of $\alpha_s(\mu_{_K})$,
$\alpha_s(\mu_{_D})$, and $\alpha_s(\mu_{_B})$,
which were determined by fitting to the measured
$2P-1S$ and $^3S_1 - ^1\!S_0$ meson splittings~\cite{Bcheat}.
    (Of course, the success of these fits is not mysterious;
there is little difficulty in accurately parametrizing a
monotonic, closely spaced set of data points
($\mu_{_B} -\mu_{_K}$ is only about 0.1 GeV) with
an adjustable fitting function which is similarly monotonic.)
    Thus the potentials $V_{\beta}(r)$, originally constructed
just to give simple scaling behaviour, actually coincide closely
with purely phenomenological potentials tuned to fit
the meson spectrum.
    This indicates that we may apply
equation (\ref{eq:BDKratios2}) with some
confidence: fitting $\Delta^{\gamma}(D)$ to
$\Delta(D)_{\rm expt} - \Delta^m(D) = 1.50 \pm 0.12$ MeV,
and taking $\beta = 1$
we obtain the results shown in columns 4 and 5 of
Table~\ref{tab:MesonResults}.
    The results are quite insensitive to
$\beta$ and to the constituent quark masses;
$\Delta^{\gamma}(B)$ and $\Delta^{\gamma}(K)$ change by
less than 5\% when beta is varied between 0 and 2,
and by less than 10\% when the quark masses are varied\cite{CQMasses}.
    We thus assign a 10\% ``parameter uncertainty error'' to
each of $\Delta^{\gamma}(B)$ and $\Delta^{\gamma}(K)$, and add it
in quadrature with the error arising from $\Delta^{\gamma}(D)$.

    Both of the predictions in Table~\ref{tab:MesonResults}
are interesting.
   First, we obtain a negative value for
$\Delta(B)$, in disagreement with most other
authors (see Table~\ref{tab:OtherDBs}).
    The sign of our prediction follows just from the
near-vanishing of $\Delta^m(B)$ (as implied by the
near-equality of the $B$ and $B_s$ hyperfine splittings),
and the fact that $\Delta^{\gamma}(B)$ is negative definite
(since the Coulomb and magnetic terms in
equation~(\ref{eq:DeltaGamma}) have the same sign).
    Second, our result for $\Delta(K)$ is in agreement
with the measured value, and even has the correct sign
(though in this case there is a large cancellation between the
quark-mass and electromagnetic terms),
so that $\Delta(K)$ can apparently be explained
without invoking large relativistic or unitarity effects.

\section{Conclusions}
\label{sec:Concl}

In summary, we have analyzed baryon and meson isospin splittings and
isomultiplet central masses in a constituent quark picture, avoiding
(as much as possible) strong reliance on particular models or
parameter sets.  Our results indicate that improved measurements of
the $\Delta$, $\Sigma_c$, $\Xi_c$, and $B^*$ splittings would be most
interesting.  The $\Delta$ masses are currently the most discrepant in
the non-charm sector, the $\Sigma_c$ and the $\Xi_c$ seem to indicate
the existence of substantial three-body effects, and measurement of
the sign of  $\Delta(B) = (B^{*-}-\bar{B}^{*0}) - (B^{-}-\bar{B}^{0})$
will provide a significant test of model calculations.

\acknowledgements
This research has been supported, in part, by an NSERC of Canada
fellowship and by the U.S. Dept. of Energy under grant
No. DE-FG02-91ER40682.

\begin{figure}
\caption{
Octet and decuplet isospin splittings (deviations from multiplet
centers) in MeV, for multiplets with $I > 0$. The fitted masses of
$j=\frac{1}{2}$ states are indicated by filled circles, and the fitted
masses of $j=\frac{3}{2}$ states are indicated by open circles. The
crosses give the experimental data and errors.  For clarity,
$j=\frac{1}{2}$ and $j=\frac{3}{2}$ points have been slightly offset.
}
\label{isospin}
\end{figure}

\begin{figure}
\caption{
Multiplet hyperfine energies, in MeV.  A spin-independent mass term
that depends on the constituent quark flavors has been subtracted from
each energy. The fitted masses of $j=\frac{1}{2}$ states are indicated
by filled circles, and the fitted masses of $j=\frac{3}{2}$ states are
indicated by open circles.  The data and errors are shown by crosses,
but only the $\Xi_c$ and $\Omega_c$ have errors of noticeable size.
}
\label{central}
\end{figure}

\begin{figure}
\caption{Hyperfine splittings in $s\bar{q}$, $c\bar{q}$, and $b\bar{q}$
mesons, as a function of the antiquark mass. (See the
text for an explanation of the $s\bar{s}$ data point.)
Typical interpolating curves are also shown.}
\label{fig:interp}
\end{figure}

\begin{figure}
\caption{The data points show effective values of $\alpha_s(\mu_{_K}),
\alpha_s(\mu_{_D}), \alpha_s(\mu_{_B})$, obtained by fitting
meson splittings to harmonic+Coulomb, linear+Coulomb, and
logarithmic+Coulomb potentials. The curves show the corresponding
behaviour of $\alpha_s(\mu)$ predicted by equation
(\protect\ref{eq:BetaHam}).}
\label{fig:AlphaS}
\end{figure}

\bigskip
\widetext
\begin{table}
\caption{Contributions to the fit to charmed baryon central masses.}
\begin{tabular}{ccccccccccccc}
    & $T_{nn}$ & $T_{ss}$ & $S_{nn}$ & $S_{ns}$
    & $A_{nc}$ & $A_{sc}$ & $D_{nc}$ & $D_{sc}$
    & total    & data     & $\pm$    & $\chi^2$ \\
\tableline
$\Lambda_c$ &0&0&213.5&0&2068.7&      0 & 0 & 0 &2282.3&2284.9&0.6 &0.053 \\
$\Sigma_c-\Lambda_c$
            &412.1&0&$-$213.5&0&0&0 &$-$27.0& 0 &171.5&167.8 &0.3 &0.047 \\
$\Xi_c$ &0&0&0&350.6&1034.4&1088.5& 0 & 0  &2473.3&2470.0&1.5 &0.130 \\
$\Omega_c$ &0&557.0&0&0&0&2176.9& 0&$-$17.6&2716.3&2719.0&7.4 &0.054 \\
$y_{c}$ &0&0&0&0&0&0&$-$6.8&6.6             & $-$0.15& 0.00 &1.60&0.008 \\
\end{tabular}
\label{tab:CharmCentral1}
\end{table}

\bigskip
\widetext
\begin{table}
\caption{Comparison of the non-charm pair terms obtained
from fits to charmed and non-charmed baryon central masses.}
\begin{tabular}{cccccc}
    & $T_{nn}$ & $T_{ns}$ & $T_{ss}$ & $S_{nn}$ & $S_{ns}$ \\
\tableline
 From non-charm data  & 412.2 $\pm$ 2.6    & 486.8 $\pm$ 3.0
 &\dec 556.8 $\pm$ 2.8 & 212.5 $\pm$ 5.4    & 351.4 $\pm$ 4.4 \\
 Including charm data & 412.1 & 486.8 & 557.0 & 213.5 & 350.6 \\
 $\chi^2$             & 0.002 & 0.000 & 0.005 & 0.034 & 0.033 \\
\end{tabular}
\label{tab:CharmCentral2}
\end{table}

\bigskip
\mediumtext
\begin{table}
\caption{Predictions for charmed baryon masses and isospin splittings.}
\begin{tabular}{cccc}
       Baryon       & Predicted mass (MeV)  & Splitting & Prediction (MeV)\\
\tableline
 $\Sigma_c^* $    & 2494 $\pm$ 16  & $\Sigma^{*1}_c$  &   1.00 $\pm$ 0.52 \\
 $\Xi_c^{\prime}$ & 2587 $\pm$ 12  & $\Sigma^{*2}_c$  &   1.64 $\pm$ 0.21 \\
 $\Xi_c^*$        & 2621 $\pm$ 15 & $\Xi_c^{\prime 1}$& $-$0.52 $\pm$ 0.33 \\
 $\Omega_c^*$     & 2743 $\pm$ 17 & $\Xi_c^{*1}$      & $-$0.44 $\pm$ 0.37 \\
\end{tabular}
\label{tab:CharmPreds}
\end{table}

\bigskip
\widetext
\begin{table}
\caption{Contributions to the fit to charmed baryon isospin splittings.}
\begin{tabular}{cccccccccc}
    & $T^1$    & $T^2$    & $S^1_s $
    & $A^1_c$  & $D^1_c$
    & total    & data     & $\pm$    & $\chi^2$ \\
\tableline
$\Sigma^1_c$ &$-$1.29& 0 &0&2.24&$-$0.11 & 0.84  & 0.90  &0.42 &0.018 \\
$\Sigma^2_c$ & 0 &1.64& 0 & 0 & 0        &1.64  &$-$1.30  &1.48&3.95 \\
$\Xi^1_c   $ & 0 &0 &$-$4.03& 1.12 &  0 &$-$2.91 &$-$6.30  &2.30&2.17 \\
\end{tabular}
\label{tab:CharmIso1}
\end{table}

\bigskip
\widetext
\begin{table}
\caption{Comparison of the non-charm pair terms obtained
from fits to charm and non-charm isospin splittings.}
\begin{tabular}{ccccc}
                     & $T^1$ & $T^2$  & $T^1_s $ & $S^1_s $ \\
\tableline
 From non-charm data &\dec $-$1.29 $\pm$ 0.0  & \dec 1.70 $\pm$ 0.21
                     &\dec $-$1.60 $\pm$ 0.26  & \dec $-$4.02 $\pm$ 0.10     \\
 Including charm data&\dec $-$1.29 &\dec 1.64  & \dec $-$1.59 & \dec $-$4.03 \\
   $\chi^2$          &\dec 0.000   &\dec 0.082 & \dec 0.000   & \dec 0.006   \\
\end{tabular}
\label{tab:CharmIso2}
\end{table}

\mediumtext
\begin{table}
\caption{Quark mass and electromagnetic contributions to
$\Delta(D)$, $\Delta(B)$, and $\Delta(K)$.}
\begin{tabular}{ccccc}
   Meson & $(dE_{hyp}/dm_q)\vert_{m_q=m_n}$ & $\Delta^m(M)$
   & $\Delta^{\gamma}(M)$ & $\Delta(M)$ \\
\tableline
$ D $ &\dec 0.004  $\;\pm$ 0.01   & \dec $-$0.02  $\pm$ 0.06
      &\dec 1.50  $\pm$ 0.12   & \dec 1.48    $\pm$ 0.10
\tablenotemark[1]\\
$ B $ &\dec 0.005  $\;\pm$ 0.01     & \dec $-$0.02    $\pm$ 0.06
      &\dec $-$0.26  $\pm$ 0.03   & \dec $-$0.29  $\pm$ 0.07\\
$ K $ &\dec $-$0.32  $\pm$ 0.06   & \dec 1.60    $\pm$ 0.44
      &\dec $-$1.75 $\pm$ 0.22   & \dec $-$0.15  $\pm$ 0.49\\
\end{tabular}
\label{tab:MesonResults}
\tablenotetext[1]{Used in fit.}
\end{table}

\bigskip
\bigskip
\narrowtext
\begin{table}
\caption{Various predictions for $\Delta(B)$. (The fourth entry is
actually a constraint obtained by fitting the measured decay constants
$f_D$ and $f_B$ to a nonrelativistic quark model.)}
\begin{tabular}{cc}
  Source   & $\Delta(B)$      \\
\tableline
 Ref. \cite{Chan}       & 0.3 $\pm$ 0.03    \\
 Ref. \cite{Sinha}      & 0.1              \\
 Ref. \cite{GodIsg}     & 0.3              \\
 Ref. \cite{Amundson}   & $-$0.05 to 0.49  \\
 This work              & $-$0.29 $\pm$ 0.07 \\
\end{tabular}
\label{tab:OtherDBs}
\end{table}


\begin{references}
\bibitem{ValenceCaveat}
Implicit here is the assumption that meson masses
depend only on their valence quark content,
{\it i.e.}, that there are no sea-quark effects.
\bibitem{CQMasses}
For the constituent quark masses, we use $m_n = 0.33$ GeV,
$m_s = 0.55$ GeV, $m_c = 1.6$ GeV, and $m_b = 5.2$ GeV.
To evaluate the stability of our results, we vary these
over the ranges $m_n = 0.30-0.36$ GeV, $m_s = 0.50-0.58$ GeV,
$m_c = 1.5-1.8$ GeV, and $m_b = 5.0-5.3$ GeV.
\bibitem{SimpleModelFlaw}
Note that the simple model described in Ref.~(2)
neglects $\langle 1/r \rangle _0 - \langle 1/r \rangle _1$.
\end{references}
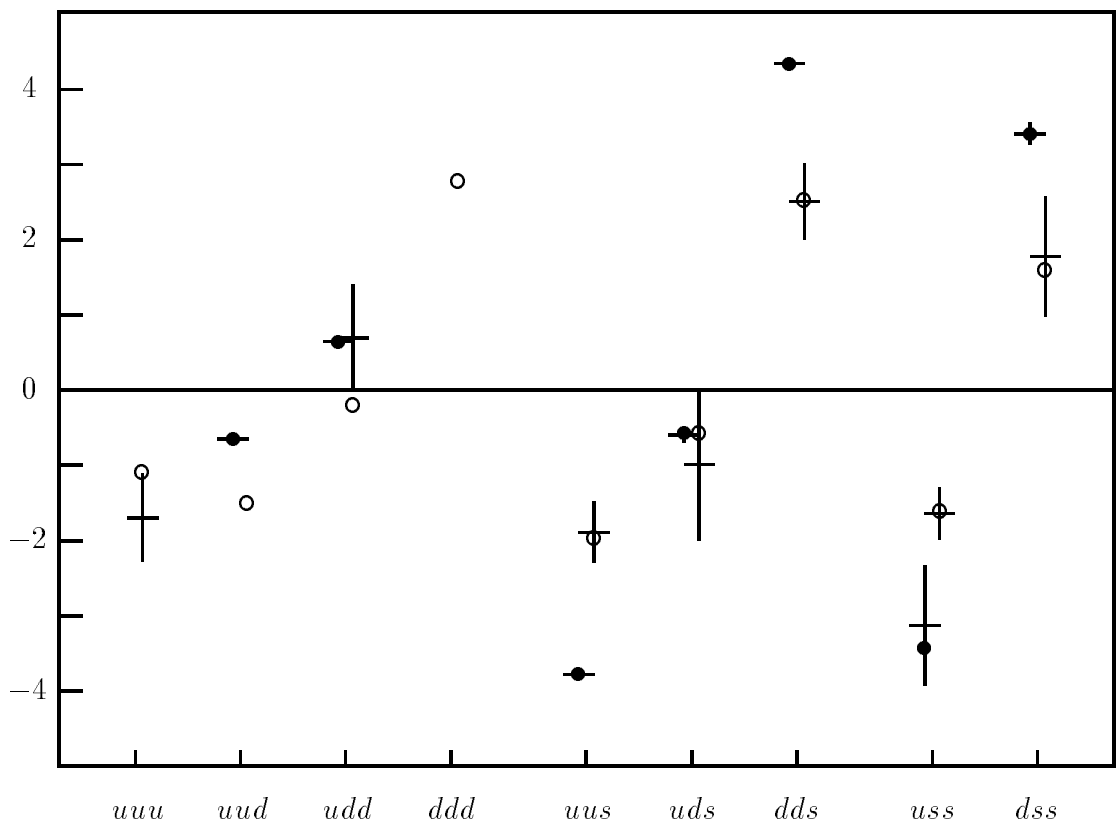
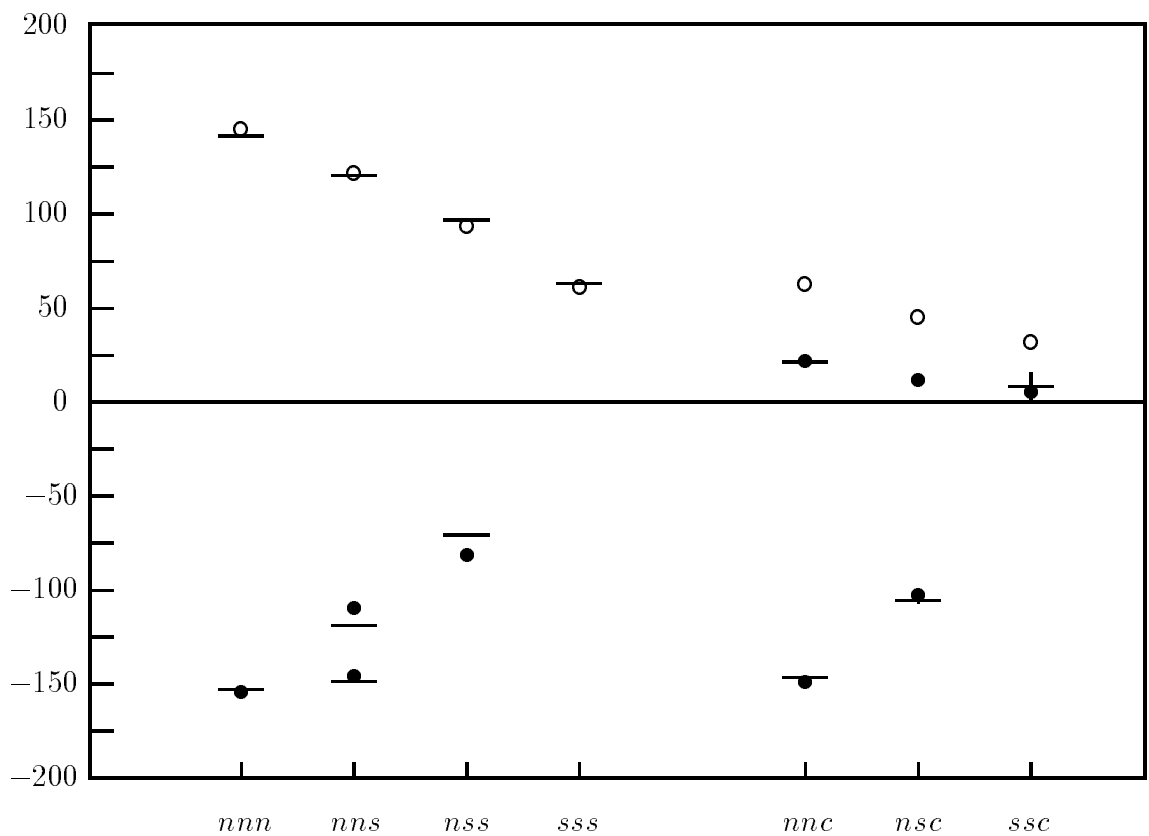
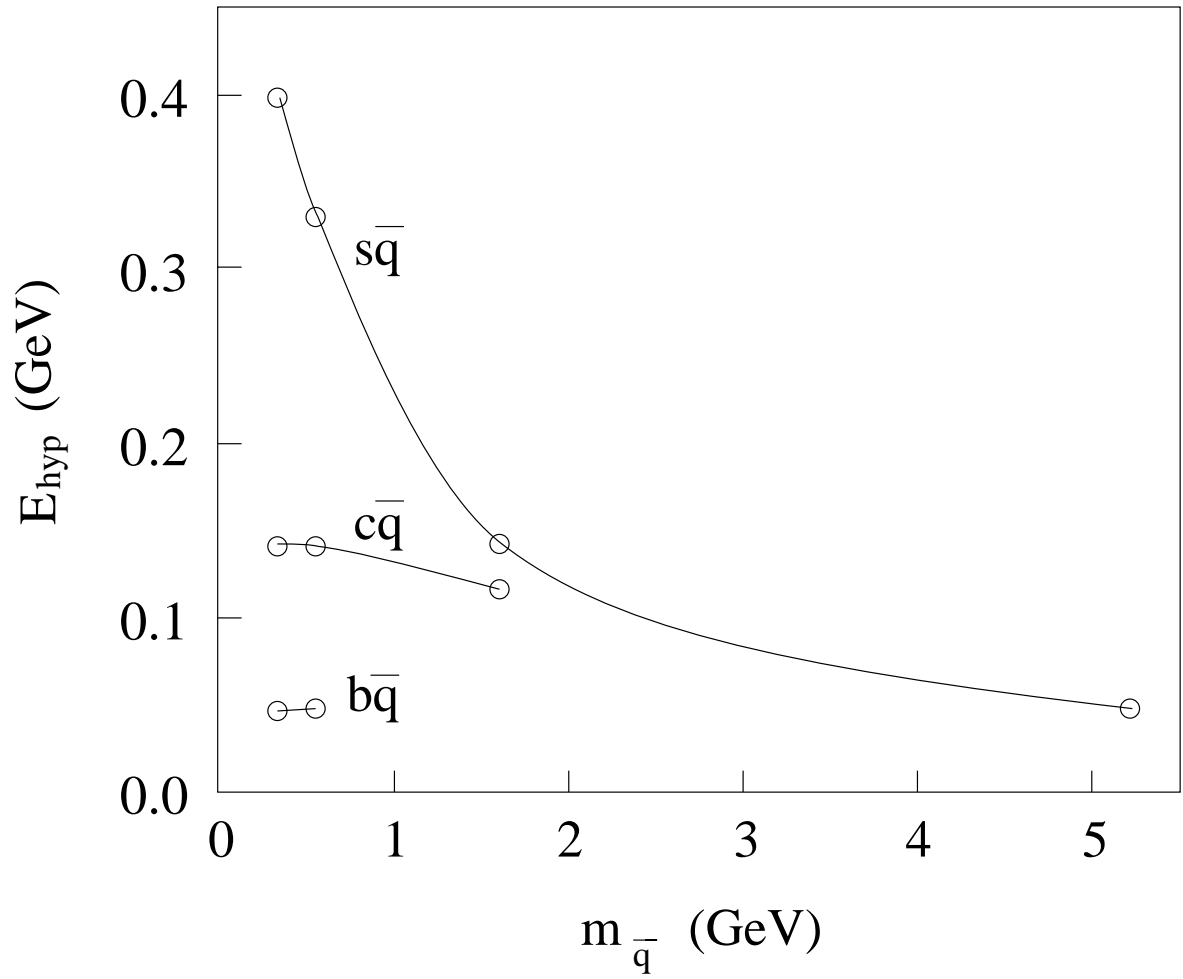
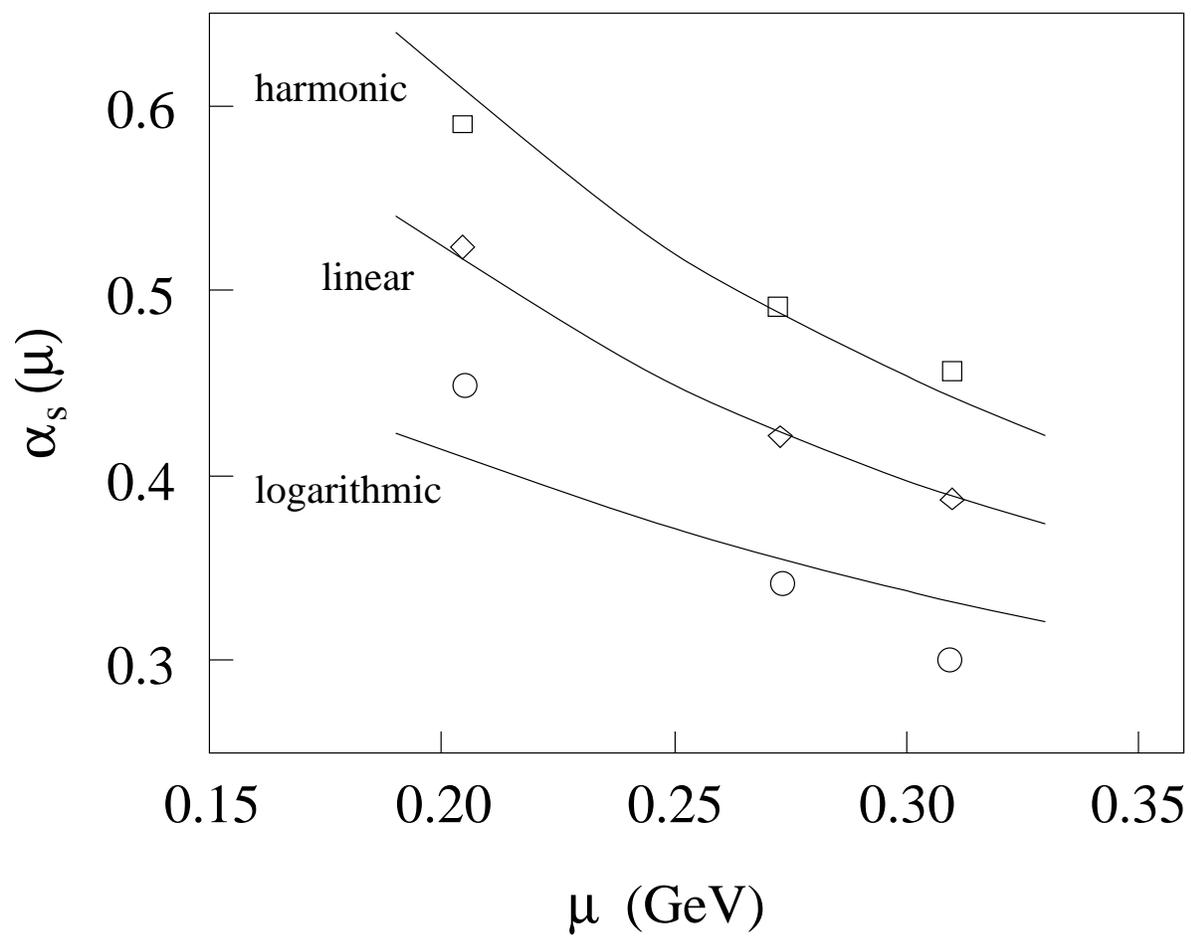
\end{document}